\documentclass[aps,prl,twocolumn,amsmath,amssymb,showpacs, superscriptaddress,notitlepage,longbibliography]{revtex4-1}

\usepackage[colorlinks=true,linkcolor=blue,anchorcolor=red,citecolor=blue, urlcolor=blue]{hyperref}
\usepackage{bm}
\usepackage{graphicx}
\usepackage{color}
\usepackage{cases}

\begin{document}

\title{Perspective: 3D quantum Hall effect}

\author{Hai-Zhou Lu}
\email{luhz@sustc.edu.cn}

\affiliation{Shenzhen Institute for Quantum Science and Engineering and Department of Physics, Southern University of Science and Technology, Shenzhen 518055, China}

\affiliation{Shenzhen Key Laboratory of Quantum Science and Engineering, Shenzhen 518055, China}

%\date{\today }
\begin{abstract}
The discovery of the quantum Hall effect in 2D systems opens the door to topological phases of matter. A quantum Hall effect in 3D is a long-sought phase of matter and has inspired many efforts and claims. In the perspective, we review our proposal that guarantees a 3D quantum Hall effect. The proposal employs the topologically-protected Fermi arcs and the ``wormhole" tunneling via the Weyl nodes in a 3D topological semimetal. The 1D edge states in this 3D quantum Hall effect show an example of ($d$-2)-dimensional boundary states. Possible signatures of the 3D quantum Hall effect have been observed in the topological Dirac semimetals, but with many questions, which will attract more works to verify the mechanism and realize the 3D quantum Hall in the future.
\end{abstract}

%\pacs{75.47.-m, 03.65.Vf, 71.90.+q, 73.43.-f}

\maketitle

\begin{widetext}

\begin{figure}[tbph]
\centering \includegraphics[width=0.9\textwidth]{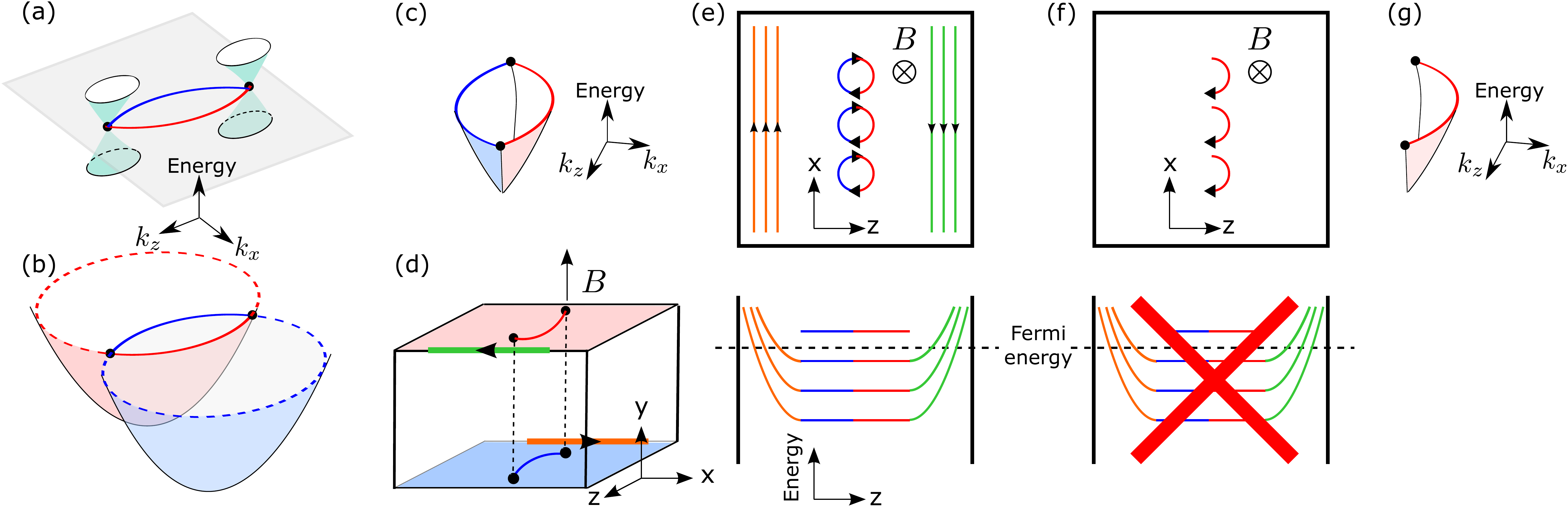}
\caption{(a) The energy dispersion of the two massless cones of 3D bulk states (green) in a topological Weyl semimetal. The black spots stand for the Weyl nodes. (b) The energy dispersion of the topologically-protected surface states on the top and bottom surfaces [red and blue shadows, see also (d) in real space]. $k_{x}$ stands for $(k_x,k_y)$ for the bulk and $k_x$ for the surface, respectively.
 The dashed curves do not exist because they are topologically forbidden, so the topological surface states look like a boat in (c), with the left and right sides from the top and bottom surfaces, respectively. (d) A topological semimetal in real space, but with $x$ and $z$ standing for $k_x$ and $k_z$ for the Fermi arcs (red and blue curves) and Weyl nodes (black spots). The green and orange arrowed lines depict the edge states of the 3D quantum Hall effect. (e) The Landau levels (red and blue) and edge states (green and orange) in the 3D quantum Hall effect, in a $y$-direction magnetic field $B$. (f) If there were only one surface, as shown in (g), an electron can not be driven by $B$ to perform a complete cyclotron motion, because it cannot take all the momentum angle from 0 to $2\pi$. As a result, there is no Landau levels, edge states, and quantum Hall effect on one surface. But two surfaces can support a complete cyclotron motion and the quantum Hall effect. The two surfaces are connected by the Weyl nodes, which are higher-dimensional singularities. According to the uncertainty principle, this ``wormhole" tunneling can connect two surfaces infinitely far apart. That is why this is called the 3D quantum Hall effect.}
\label{Fig:3DQHE}
\end{figure}

\end{widetext}

In a magnetic field, a moving charge feels a Lorentz force orthogonal to both its velocity and the magnetic field, leading to the Hall effect. Klaus von Klitzing discovered that in strong magnetic fields the Hall resistance of a 2D electron gas can be quantized into a series of plateaus in terms of $(h/e^2)/n$ \cite{Klitzing80prl}, where $e$ is the elementary charge, $h$ is Planck¡¯s constant, $n$ is an integer known as ¡°Chern number¡± (named after mathematician Shiing-Shen Chern). The quantum Hall effect has led to three Nobel Prizes in physics (1985 von Klitzing; 1998 Tsui, Stormer, Laughlin; 2016 Thouless, Haldane, Kosterlitz).
Usually, the quantum Hall effect takes place only in 2D systems.
In a strong magnetic field, the energy spectrum of a 2D electron gas is quantized into Landau levels. The Landau levels deform at the sample edges and cross the Fermi energy, forming 1D edge states.
Electrons can flow through the edge states dissipationlessly.
When the Fermi energy is placed between two Landau levels, each edge state contributes a Hall conductance of $e^2/h$ and vanishing longitudinal conductance in the Hall-bar measurement.
The quantization can be observed in two dimensions because the bulk states in the interior of the sample can be gapped.
In contrast, a magnetic field quantizes the energy spectrum of a 3D electron gas into 1D Landau bands that disperse along the direction of magnetic field. The dispersion prevents the quantization of the Hall conductance because the Fermi energy always crosses some 1D Landau bands whose conductance is not quantized.Different schemes have been proposed to gap the 3D bulk states for the quantization of the Hall conductivity in three dimensions \cite{Halperin87jjap,Kohmoto92prb,Koshino01prl,Bernevig07prl,Jin18arXiv}. Nevertheless, a 3D quantum Hall effect remains a long-sought phase of matter \cite{Stormer86prl,Cooper89prl,Hannahs89prl,Hill98prb,Cao12prl,Liu16nc,Masuda16sa,Tang18arXiv}.

We propose a 3D quantum Hall effect with a quantized Hall conductance in a topological semimetal \cite{WangCM17prl}.
The band structure of a topological semimetal looks like a 3D graphene \cite{Wan11prb,Yang11prb,Burkov11prl,Xu11prl}, with the conduction and valence bands touching at the Weyl nodes [Fig \ref{Fig:3DQHE} (a)]. For momenta $k_z$ between the Weyl nodes, this band structure is equivalent to a 2D topological insulator, with topologically protected states on the surfaces [Fig. \ref{Fig:3DQHE} (b)-(d)] parallel to the $z$ direction. The Fermi surface of the surface states is known as the Fermi arcs [red and blue curves in Figs. \ref{Fig:3DQHE} (a)-(d)].
The Fermi-arc surface states form a unique 2D electron gas, half from the top surface and half from the bottom surface [Figs. \ref{Fig:3DQHE}(c) and (d)]. It may host a quantum Hall effect. If there were only the top surface [Fig. \ref{Fig:3DQHE}(g)], the Fermi-arc surface states cannot support a complete cyclotron motion in real space [Fig. \ref{Fig:3DQHE}(f)], then there is no Landau levels, edge states, and the quantum Hall effect. Fortunately, the top and bottom surfaces can form a complete 2D electron gas, with a closed Fermi surface connected by the Weyl nodes. Driven by the $y$-direction magnetic field, an electron performs half of an cyclotron motion on the top Fermi arc, then tunnels via a Weyl node to the bottom Fermi arc to completes the cyclotron motion. In this way, the top and bottom Fermi arcs together support a complete cyclotron motion and the quantum Hall effect. More importantly, the Weyl nodes are 3D singularities in momentum space, so according to the uncertainty principle they can connect the 2D surfaces separated far apart infinitely in real space. That is why we call it a 3D quantum Hall effect. This is like the wormhole effect, which connects 3D spaces via higher-dimensional singularities.

Recently, the quantized Hall resistance plateaus have been experimentally observed in the topological semimetal Cd$_3$As$_2$ \cite{Uchida17nc,ZhangC17nc-QHE,Schumann18prl}, with thickness ranging from 10 to 80 nm. They cannot be regarded as 2D.
Nevertheless, several questions still hold. First, Cd$_3$As$_2$ is a Dirac semimetal, composed of two time-reversed Weyl semimetals. At a single surface, there is a complete 2D electron gas, formed by two time-reversed half 2D electron gases of the Fermi-arc surface states. There may be also the trivial quantum Hall effect on a single surface. Second, the 3D bulk states quantize 2D subbands for those thicknesses. If the 3D bulk states cannot be depleted entirely, they also have the trivial quantum Hall effect. The two issues may explain the 2-fold and 4-fold degenerate Hall resistance plateaus observed in the experiments. To deplete the 3D bulk states, the Fermi energy has to be placed exactly at the Weyl nodes.
How to distinguish these trivial mechanisms from the 3D quantum Hall effect will be an interesting direction. Previously, when studying the geometric phase, the parameter space is usually either in real space or momentum space \cite{Xiao10rmp}. The Weyl orbit formed by the Fermi arcs and Weyl nodes is a new physics, because part of the geometric phase is accumulated in real space and part in momentum space, quite different from the parameter spaces studied before. In particular, the geometric phase has a thickness dependence when accumulated along the path as electrons tunnel between the top and bottom surfaces \cite{Potter14nc,ZhangY16srep}. Recently, a new experiment uses this thickness-dependent phase shift to demonstrate the contribution of the Weyl orbit in the observed quantized Hall resistance \cite{ZhangC18nat}. 

In this 3D quantum Hall effect, the edge states are located at only one edge on the top surface and at the opposite edge on the bottom surface [green and orange arrowed lines in Figs. \ref{Fig:3DQHE}(d) and(e)], which may be probed by scanning tunneling microscopy \cite{ZhengH16acsnano} or microwave impedance microscopy \cite{Ma15nc}. The 3D quantum Hall effect may be realized in other systems with novel surface states. More works will be inspired to verify the mechanism and realize the 3D quantum Hall effect in the future.

This work is supported by Guangdong Innovative and Entrepreneurial Research Team Program (2016ZT06D348), the National Key R \& D Program (2016YFA0301700), the
National Natural Science Foundation of China (11574127), and the Science, Technology, and Innovation Commission of Shenzhen Municipality (ZDSYS20170303165926217, JCYJ20170412152620376)).

%\bibliographystyle{apsrev4-1-etal-title}
%\bibliography{refs-transport,ref-3dqhe,refs-oscillation,refs-type2Exp,refs-majorana,refs-PHE}


\end{document}